# A many-to-many assignment game and stable outcome algorithm to evaluate collaborative Mobility-as-a-Service platforms

**Theodoros P. Pantelidis [1], Joseph Y. J. Chow [*1], Saeid Rasulkhani [2]**
[1]C2SMART University Transportation Center, New York University Tandon School of Engineering, Brooklyn, NY, USA
[2]Scoop Technologies, San Francisco, CA, USA
*Corresponding author: joseph.chow@nyu.edu

## ABSTRACT
As Mobility as a Service (MaaS) systems become increasingly popular, travel is changing from unimodal trips to personalized services offered by a platform of mobility operators. Evaluation of MaaS platforms depends on modeling both user route decisions as well as operator service and pricing decisions. We adopt a new paradigm for traffic assignment in a MaaS network of multiple operators using the concept of stable matching to allocate costs and determine prices offered by operators corresponding to user route choices and operator service choices without resorting to nonconvex bilevel programming formulations. Unlike our prior work, the proposed model allows travelers to make multimodal, multi-operator trips, resulting in stable cost allocations between competing network operators to provide MaaS for users. An algorithm is proposed to efficiently generate stability conditions for the stable outcome model. Extensive computational experiments demonstrate the use of the model to handling pricing responses of MaaS operators in technological and capacity changes, government acquisition, consolidation, and firm entry, using the classic Sioux Falls network. The proposed algorithm replicates the same stability conditions as explicit path enumeration while taking only 17 seconds compared to explicit path enumeration timing out over 2 hours.

## 1. INTRODUCTION

There is a growing need to focus on managing the capacities, allocation, and pricing of mobility services in a Mobility-as-a-Service (MaaS) (Hensher, 2017; Djavadian and Chow, 2017) ecosystem. Under this ecosystem, city agencies play a key role as facilitators in either economic deregulation through relationships with suppliers, or through government contracting with the operators, as illustrated by the evolution from **Figure 1a** to **1b** or **1c** (Wong et al., 2019). As such, city agencies need to be able to assess the impact on other mobility operators and travelers when a new mobility operator enters the market, or an existing one changes their service capacity, routing algorithm, or pricing mechanism. A new mobility service or change to an existing one can cause travelers to switch routes or combine the service of one operator with those of other operators to fulfill their trips. It can lead to certain routes becoming unstable to operate. Changes in algorithms (e.g. Stiglic et al., 2015) or government policies like ride surcharges (e.g. Hu, 2019) can alter the allocation of costs between travelers and operators. Any MaaS market equilibrium model developed for the public policymaker needs to be sensitive to both traveler (multimodal/multi-operator routes) and operator (service coverage, fleet, pricing) decisions.

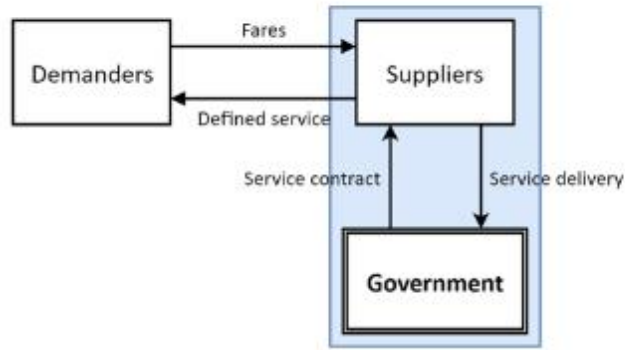

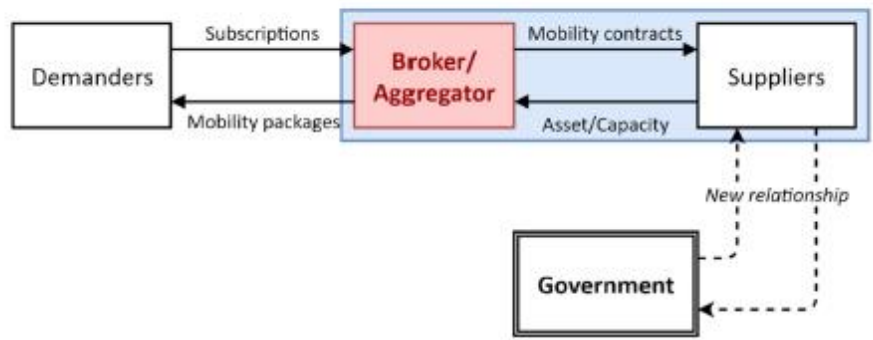

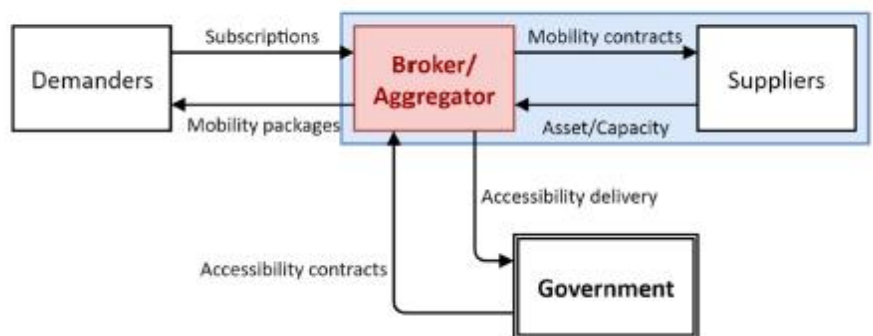

**Figure 1**. Evolution from current service delivery model (A) to two alternative models under MaaS (B, C) (source: Wong et al., 2019).

For this purpose, classic traffic assignment models that emphasize only traveler route decision-making are not effective tools. In a MaaS setting the policy questions are not focused on congestion on the roadway, but instead on how travelers match to different combinations of fixed route public transit and/or various mobility options (e.g. bikeshare, ride-hail, microtransit, among others) considering capacities of these systems and their cost allocation policies (e.g. fares transfer dollars to service, stop locations trade-off between access time). For capacity we consider effective planning-level service capacities; i.e. a bike share service that uses a certain rebalancing algorithm would provide a certain maximum flow from one location to another.

Rasulkhani and Chow (2019) proposed such a method for evaluating unimodal trip systems, where each operator acts as a set of service routes and each traveler matches "many-to-one" to one route while ensuring the line capacities are not violated and stability conditions from the core (Shapley and Shubik, 1971) are met. The model is computationally tractable and can be solved using classic algorithms for capacitated assignment and linear programming (for the stable outcome problem). Unlike network flow games (e.g. Bird, 1976; Derks and Tijs, 1985; Fragnelli et al., 2001; Agarwal and Ergun, 2008) that form coalitions between operators, the model matches between travelers and operators so that it explicitly captures both operator and travel behavior in a network of mobility markets. The model from Rasulkhani and Chow (2019) does not handle matching of travelers' paths to multiple operators for modeling MaaS platforms. Use cases for such models abound: city agencies may act as cyber-physical platform providers in which mobility operators and travelers match with transaction fees that depend on the design of the built environment (see Chow, 2018). MaaS policy-makers, including government agencies and transport providers, can then use the solution of such a model to make trade-offs for designing their platform under different cost allocation policies and algorithms, link capacities, and determine negotiating power of different operators for the purpose of forming coalitions or justifying subsidies between operators to match with traveler paths. We illustrate these use cases that need to be modeled in **Table 1**.

**Table 1**. Sample use cases for a planning model for a public-operated MaaS platform

| Use case | Model parameters | Required model output |
|---|---|---|
| *Technology*: evaluate/regulate market due to new algorithm or operating policy from an operator | Changes to travel disutilities of travelers (which may be in-vehicle, access, or wait time), link operating costs, or link capacities of operators | Impact on operator-routes that stay in market, passenger link flows, and how their stable price range changes |
| *Subsidy*: platform may subsidize one or more of the operators | Change in threshold for an operator to leave a market (they might be able to operate at a loss up to a threshold); cost allocation for that operator may align with welfare maximizing instead of profit maximizing | Links that can be operated in this setting, revenues and flows under the changed setting |
| *Tax*: platform may impose a surcharge on a subset of operators | Changes to operating cost for the operators | Changes in operating links, flows, and shifts in stable pricing range as a result of surcharge |
| *Merger*: two or more operators in a platform may merge or ally | The stability conditions would treat those operators as a single operator | Changes in revenue and ridership due to the merger |
| *Investment*: evaluate/regulate market due to increased investment by an operator on their fleet size, new service coverage area, etc. | New candidate links/nodes in network, changes to link capacities | Whether those links stay in the market, subsequent flows, prices for new services as well as impacts on other operators |
| *Disruption*: links may be closed or degraded | Closure of links/nodes in network, changes to link capacities | Whether those links stay in the market, subsequent flows, prices for new services as well as impacts on other operators |

The challenge of matching multiple links of different traveler paths to multiple operators is a many-to-many assignment game. In such a game, the stability conditions become more complex because they need to be considered from both a user's path level as well as an operator's level in serving that user. In a MaaS market, each operator owns one or more links and may choose not to serve that link (i.e. exit the market) if there is insufficient incentive to provide service there. Each link is capacitated with planning-level service capacities. Furthermore, each operator can choose a price to charge for using their link to the users. How should competing operators sharing different legs of a traveler's trip set their prices? Blocking pairs" may form to prevent a stable path from forming, which are not trivial to model in this setting.

We propose a model for this many-to-many assignment game and show how to derive an optimal assignment flow and corresponding stable outcome space between the operators and the travelers or users. If a stable outcome space exists, it provides boundaries over which a city agency can work with competing operators to allocate costs between users and each operator to set prices, as illustrated in the use cases in **Table 1**. The model does not assume any specific cost allocation mechanism or policy used by each operator; it only determines the thresholds within which such mechanisms are stable. An empty stable outcome space implies the unsustainability of the platform as designed, which would warrant further planning (e.g. changing travel costs through infrastructure investments or policies, adding further capacities, or introducing additional candidate routes for operators to serve).

## 2. LITERATURE REVIEW

While several studies have examined demand for MaaS services (Strömberg et al., 2018; Matyas and Kamargianni, 2019a,b), these studies have not sought to quantify or structure the relationships between decisions made by operators and users in providing and consuming routes in a MaaS market. Earlier research on coexisting operators (see Chow and Sayarshad, 2014, for a review) consider noncooperative games, including a generalized Nash equilibrium for a duopolistic market of private mass transit operators (Harker, 1988) and toll pricing operators (Yang and Woo, 2000; Zhang et al., 2011). Cooperative game research, on the other hand, determines the cost allocations necessary to support cooperation as a coalition formation problem. Examples of cooperative games include Agarwal and Ergun (2008) and Lu and Quadrifoglio (2019), as well as the network flow games cited earlier. Cooperative games within a transportation environment to support multimodal trips are called *collaborative transportation problems*.

Collaborative transportation problems have been studied (Schulte *et al.*, 2019) primarily in truck and airline scheduling. Furthermore, profit allocations for collaborative transportation systems have been studied (Algaba *et al.*, 2019). However, the models are from the perspective of the operators as specific allocations are assumed using simple mechanisms such as equalitarian and proportional profit divisions between operators. We study collaborative transportation for MaaS from the perspective of the platform provider, where stable core allocations are obtained by solving a mathematical program that incorporates stability conditions.

The basis for this latter type of model is a stable matching model that forms coalitions between users and operators that no user or operator has incentive to break. Stable matching models in which utilities are transferable (TU-games) between buyers and sellers are called assignment games, first formulated as linear programs (LPs) by Shapley and Shubik (1971).

In the basic model from Shapley and Shubik (1971), there is a set of buyers $P$ and sellers $Q$. A buyer $i \in P$ that matches with a seller $j \in Q$ providing the product at cost $c_j$ earns a utility of $U_{ij}$. The difference between the utility and cost of production is the payoff $a_{ij} = \max(0, U_{ij} - c_j)$. A successful match means the seller transfers the utility to the buyer with a price $p$. The buyer earns utility equal to $u_i = U_{ij} - p$ while the seller profits $v_j = p - c_j$. The basic assignment game is shown in Eq. (1).

$$\max \sum_{i \in P} \sum_{j \in Q} a_{ij} x_{ij} \tag{1a}$$

Subject to

$$\sum_{i \in P} x_{ij} \leq q_j, \quad \forall j \in Q \tag{1b}$$

$$\sum_{j \in Q} x_{ij} \leq w_i, \quad \forall i \in P \tag{1c}$$

$$x_{ij} \in \{0,1\}, \quad \forall j \in Q, i \in P \tag{1d}$$

where $x_{ij}$ is a binary variable whether a match occurs, and $q_j$ and $w_i$ are quotas for each side. If $q_j$ and $w_i$ are equal to one, the assignment game is one-to-one. Nonsingular integers reflect many-to-one or many-to-many games. An outcome $((u,v); x)$ of the game is feasible if $u_i \geq 0$ and $v_j \geq 0$ and satisfies the constraints (1b) – (1d) as well as $\sum_{i \in P} u_i + \sum_{j \in Q} v_j = \sum_{i \in P} \sum_{j \in Q} a_{ij} x_{ij}$. A feasible payoff is stable if $u_i + v_j \geq a_{ij}$ when $x_{ij} = 0$. The core of the assignment game corresponds to the solutions of the dual of the LP.

In the multiple partner assignment game (Sotomayor, 1992), the values to the buyers and sellers are distributed to different matching partners: $u_{ij}$ is the utility gained by buyer $i$ when matched to seller $j$ and $v_{ij}$ is the profit gained by seller $j$ when matched to buyer $i$. A feasible outcome $((u,v); x)$ is stable if $u_i + v_j \geq a_{ij}$ when $x_{ij} = 0$, where $\min_j \{u_{ij}\} \geq 0$ and $\min_i \{v_{ij}\} \geq 0$. Under these games, two extreme vertices of the stable outcome space can be identified as the "buyer optimal" and the "seller optimal" outcomes reflecting ideal outcomes for each side between which cost allocation mechanisms can be negotiated. If the outcome space is empty, it means the optimal assignment is not stable.

Several studies have been conducted using stable matching (e.g. Cseh and Skutella, 2014; Wang et al., 2017; Lin et al., 2018; Peng et al., 2018; Zhang and Zhao, 2018; Lu and Quadrifoglio, 2019; Yang et al., 2019) generally as a mechanism for optimizing ridesharing services, in a normative sense. These have not considered evaluation of a city platform design by modeling a market equilibrium for a MaaS system of multiple operators.

Rasulkhani and Chow (2019) proposed such a descriptive assignment game model where buyers are traveler OD pairs and sellers are bundles of service routes. In this case, each operator $f \in F$ owns a set of one or more routes $R_f$, where $R = \bigcup_{f \in F} R_f$. Each service route $r$ consists of a set of links $A_r$ owned by an operator; one operator may own multiple service routes. The resulting many-to-one assignment game between multiple user OD pairs and different links of each service route is characterized by Eq. (2) for a set of users $S$, demand $d_s$, and a service route capacity $w_r$ defined as the load that cannot be exceeded anywhere along a route. The index $\{k\}$ is used to refer to a dummy user that is matched to unused routes. Eq. (2d) ensures that routes that are unused are matched to the dummy user. The parameter $\delta_{asr}$ is set to 1 if a match between user $s$ and route $r$ uses link $a$ and 0 otherwise. $M$ is a big constant. $a_{sr} = \max(0, U_{sr} - t_{sr})$, where $t_{sr}$ is the generalized travel disutility for user $s$ matched to route $r$.

$$\max \sum_{s \in S} \sum_{r \in R} a_{sr} x_{sr} \tag{2a}$$

Subject to

$$\sum_{r \in R} x_{sr} \leq d_s, \quad \forall s \in S \setminus \{k\} \tag{2b}$$

$$\sum_{s \in S \setminus \{k\}} \delta_{asr} x_{sr} \leq w_r, \quad \forall a \in A_r, r \in R \tag{2c}$$

$$\sum_{s \in S \setminus \{k\}} x_{sr} \leq M(1 - x_{kr}), \quad \forall r \in R \tag{2d}$$

$$x_{sr} \in \{0, \mathbb{Z}_+\}, \quad \forall s \in S \setminus \{k\}, r \in R \tag{2e}$$

$$x_{kr} \in \{0,1\}, \quad \forall r \in R \tag{2f}$$

A fare $p_{sr}$ is charged to each member of user group $s$ for matching to route $r$. The operating cost of a route $r$ is set to $C_r$ and is distributed to each user as $c_{sr}$. $B(r,x)$ $(B(s,x))$ is the set of users (routes) matched to route $r$ (user $s$) in assignment $x$. $\bar{R}$ is the set of routes matched to at least one user. $G_r$ is the set of user groups that can be feasibly matched to route $r$. $v_r = \sum_{s \in C(r,x)} v_{rs}$ is the total benefit that route $r$ gains from matches in assignment $x$. The stable outcome space corresponding to this assignment game is defined by the following set of constraints in Eq. (3), where Eq. (3a) – (3d) represent the feasibility conditions and Eq. (3e) is the stability condition.

$$\sum_{s \in B(r,x)} u_s + v_r = \sum_{s \in B(r,x)} a_{sr} - C_r, \quad \forall r \in \bar{R} \tag{3a}$$

$$v_r = 0, \quad \forall r \in R \setminus \bar{R} \tag{3b}$$

$$u_s = 0, \quad \forall s: B(s,x) = \emptyset \tag{3c}$$

$$u_s \geq 0, v_{sr} \geq 0, \quad \forall r \in \bar{R}, s \in S \tag{3d}$$

$$\sum_{s \in G_{r'}} u_s + v_r \geq \sum_{s \in G_{r'}} a_{sr} - C_r, \qquad \forall G_{r'}: r' \notin R_f, r \in R_f, f \in F \tag{3e}$$

The buyer-optimal and seller-optimal vertices of the space can be found by maximizing $Z = \sum_{s \in S} u_s$ (buyer-optimal) or $Z = \sum_{r \in R} v_r$ (seller-optimal) as LPs, and any solution within the space can be interpolated since the space is convex. Specific cost allocation mechanisms can also be sought by setting the appropriate objective $Z$. While different mechanism designs can be incorporated, the scope of this work is on defining the stable outcome space, so a comprehensive review of such studies is not provided. Readers are referred to Rasulkhani and Chow (2019) instead.

As shown in the formulations, each user $s$ is treated as an OD pair as opposed to a path of multiple operator routes with transfers. This would therefore not model a MaaS setting. The assignment game needs to be redefined from a link-based perspective, where each operator owns a set of links (each link connecting two stops, zone centroids, or transfer points) and users are now distributed over different paths consisting of links involving transfers from one collaborative operator to another.

## 3. PROPOSED METHODOLOGY

**Problem notation**

$F$: set of operators, where $f = 0$ is a dummy operator representing no operator in the platform
$N$: set of nodes in the platform
$A$: set of links in the network
$A_f$: a disjoint subset of $A$ owned by operator $f \in F$
$N_i(+), N_i(-)$: the set of heads (+) and tails (-) that are formed by links connected to node $i \in N$
$S$: set of traveler groups represented by OD from one node to another
$O(s), D(s)$: origin-destination nodes of $s \in S$
$R_s$: set of feasible paths that can serve OD $s \in S$
$R_s^*$: set of optimal paths for which there may be flow for OD $s \in S$
$A_r$: set of links that form path $r \in R_s$
$R_f$: set of routes in which operator $f \in F$ operates one or more links
$x_{ij}^s$: flow variable for link $(i,j) \in A$ and OD $s \in S$
$t_{ij}$: generalized link travel cost $(i,j) \in A$, in units of $
$c_{ij}$: link operating cost $(i,j) \in A$, in the same units of $
$y_{ij}$: binary variable indicates if link $(i,j) \in A$ has flow
$d_s$: number of homogeneous travelers for OD $s \in S$
$w_{ij}$: capacity of link $(i,j) \in A$

$U_s$: utility of $s \in S$, in units of \$

$u_s$: consumer surplus $s \in S$, in units of \$

$p_{rf}$: price of operator $f \in F$ on path $r \in R_s$, in units of \$

$z_r^*$: flow on path $r \in R_s$

$\mu_{ij}^*$: capacity dual variable of $(i,j) \in A$, in units of \$

$\gamma_{ij}$: a link-level subsidy that effectively reduces the cost of operation for link $(i,j) \in A$, in units of \$

$\delta_{rf}$: binary parameter to indicate whether operator $f \in F$ is on path $r \in R_s$

### 3.1. Model formulation

We present a many-to-many assignment game model to be used to evaluate MaaS networks for planning and evaluation of public agency platforms. We define two disjoint sets of players: the first set includes all network users $s \in S$ that represent distinct O-D demand pairs and the second set includes all the operators $f \in F$ that provide transportation services on network links. Some links may not be owned by any operator; we assume these are owned by the platform. Examples include transfer links between two operators, which may involve walking or waiting in between services. Those links are owned by a dummy operator $f = 0$.

In this assignment users are matched to a feasible user path $r$ from an origin node $O(s)$ to a destination node $D(s)$ by being served by a set of operators $F_r$ for the links forming that path. In a capacitated network, a user group $s$ may be distributed over a path set $R_s^*$ among a feasible set $R_s$. The network is defined as $G(N,A)$ where $G$ is composed of a set of mutually exclusive operator-owned subgraphs $G_f$, $f \in F$, and $G = \bigcup_{f \in F} G_f$. Let $N_i(+)$ and $N_i(-)$ respectively be the sets of heads and tails formed from links connected to node $i \in N$.

The output for a given set of operator service links and users along with operating costs $c_{ij}$ and link capacities $w_{ij}$ for each link $(i,j) \in A$ and user travel disutilities $t_{ij}$ is a set of link flows $x_{ij}^s$ per user group $s$, identity of unmatched operator links, and corresponding stable outcome $((u,v),x)$ space for user value $u$, operator profits $v$, and fares $p_{rf}$ paid by each user-path $r$ to each operator $f$. In the case where a fixed fare is assumed, one can add an additional constraint to require all $p_{rf} \equiv p_f$ (which reduces the feasible outcome space). The model assumes costs are transferable between users and operators, so the operating costs, fares, and travel disutilities should all be in a common monetary unit (\$). The fare represents the generalized cost allocation from user to operator, and can represent not only monetary fare, but also an increase of user access cost to the operator, for example.

The behavioral interpretation of this model structure is the same as that of the assignment game from Shapley and Shubik (1971): the matching problem identifies the optimal assignment between multiple operators and multiple travelers while the behavioral incentives of the travelers and operators are determined in the stable outcome subproblem to ensure they would match only if there's a benefit to do so. Other model structures for markets of operators and travelers have been proposed in the past. For example, centralized operators with travelers have been modeled as bilevel optimization problems (see Yang and Bell, 1998), and markets of multiple operators as a generalized Nash equilibrium with a bilevel structure for travelers (Zhou et al., 2005). A bilevel optimization suggests a leader-follower relationship that assumes an agency makes the decisions first. This makes sense in the case where the agency is making capital expenditure decisions that can take years to complete. In the case of a MaaS platform, services are decided in a much shorter, tactical planning level time horizon (most MaaS markets today did not exist five years ago), and in competitive response from one operator to another. By the same token, it would not make sense for platforms like Airbnb to model its sellers as leaders and its buyers as followers, which is why models for those follow a two-sided market structure instead (see Rochet and Tirole, 2003).

Shapley and Shubik (1971) showed the assignments are obtained from the primal matching problem and the stable outcomes are determined from the dual problem forming the core. Rasulkhani and Chow (2019) proposed an integer program for the assignment and a linear program that captured the stability conditions of the integer program. We similarly propose an integer program for the flow assignment and a linear program for the corresponding stable cost allocations of user value and operator profit.

### 3.1.1. Matching problem

We propose to formulate the matching problem using a familiar formulation from network design (Magnanti and Wong, 1984; Gendron and Larose, 2014): the multicommodity capacitated fixed-charge network design problem (MCND). Each commodity corresponds to user group $s$. Each user group $s$ is characterized by demand $d_s$ for that unique O-D pair. A binary variable $y_{ij} = 1$ if link $(i,j)$ is operated by its owner and 0 otherwise. Since we seek the optimal flow assignment, the identification of operators for the links are not needed at this stage.

$$\phi(N) = \min \sum_{(i,j) \in A} \sum_{s \in S} t_{ij} x_{ij}^s + \sum_{(i,j) \in A} c_{ij} y_{ij} \tag{4a}$$

Subject to

$$\sum_{j \in N_i(+)} x_{ij}^s - \sum_{j \in N_i(-)} x_{ji}^s = \begin{cases} d_s, & if\ i = O(s) \\ -d_s, & if\ i = D(s), \\ 0 & otherwise \end{cases} \quad \forall i \in N, s \in S \tag{4b}$$

$$\sum_{s \in S} x_{ij}^s \leq w_{ij} y_{ij}, \quad \forall (i,j) \in A \tag{4c}$$

$$x_{ij}^s \geq 0, \quad \forall (i,j) \in A, s \in S \tag{4d}$$

$$y_{ij} \in \{0,1\}, \quad \forall (i,j) \in A \tag{4e}$$

Objective function (4a) minimizes total costs which includes the generalized travel costs of passenger and operation costs of operators. Constraint (4b) ensures the feasibility of flow in the network. Constraint (4c) is the capacity constraint for each link and constraints (4d) and (4e) are the non-negativity and integral constraints. The model is an extension of the original assignment game from Shapley and Shubik (1971) to a MaaS setting, where travelers between OD pairs are the buyers and operator-links are the sellers. The many-to-many matching occurs through the formation of user path flows. The relationships between the matching sets are shown in **Figure 2**, where the basic buyer-to-seller matching is extended to be between many paths of users to many links of operators.

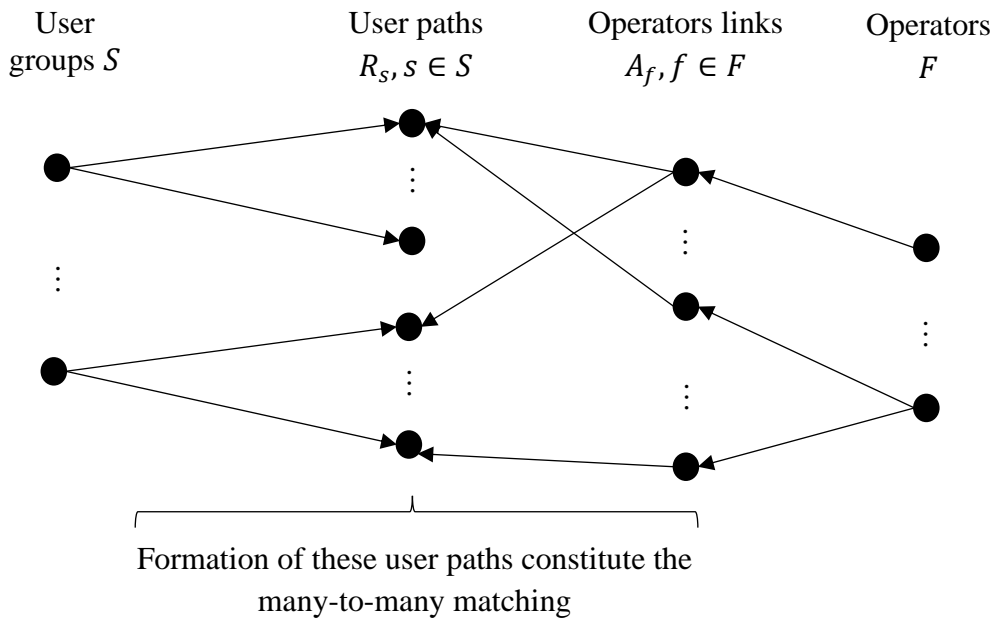

Figure 2. Illustration of the many-to-many matching structure in the MaaS market model.

The model as formulated assumes all demand is assigned onto the network. Elastic demand can be modeled. As noted earlier, links assigned to a dummy operator $f = 0$ are used for transfers by noting that stability conditions do not need to be applied to $f = 0$. These links can also be used to connect an OD $s$ directly to represent all alternative mobility options outside of the platform: competing platforms or operators, staying at home, or having the travelers go by personal car instead of using a mobility service. The maximum utility of all those options can be represented by subtracting that from the utility $U_s$ and applying that difference as the link cost. If there are no other options available, the link cost can be set to $U_s$.

Unlike the conventional use of the MCND, which is to find a subset of new links within a budget to build out, this use of the model determines which operator-links should enter the market under optimal market matching. For example, conventional use of MCND assumes some existing network upon which a subset of new candidate links is being considered. In our use of the model, *all links* are treated as candidate links and are each owned by one operator.

By casting the matching problem in a MCND structure, it can be solved using conventional MCND solution methods like branch-and-bound-and-cut algorithms (Gendron and Larose, 2014).

### 3.1.2. Stable outcome problem

The more novel contribution of this study is to propose a corresponding stable outcome model to the MCND to establish the stability conditions for the operators that choose to stay in the market. The binary parameter $\delta_{ijr}$ indicates whether a user path $r \in R$ is incident on link $(i,j) \in A$. The link set $A_r \subseteq A$ consists of links included in path $r \in R_s$, while $A_f \subseteq A$ consists of links owned by operator $f \in F$. The set of links owned by operator $f \in F$ along path $r \in R_s$ is formed by the intersection $A_f \cap A_r$. Each user $s$ generates payoff (trip utility) $U_s$ to realize their trip. Let $R_f$ be the set of routes in which operator $f$ serves.

The constraints of the corresponding stable outcome problem are derived as follows. For each user group $s \in S$ the core payoff is denoted as $u_s$, the utility surplus from making a trip. For each operator $f \in F$, a price $p_{rf}$ is charged once per path. The operator's payoff is the total revenue collected from network users $\sum_{r \in R_f} p_{rf} z_r$ based on the price $p_{rf}$ charged to user-path $r$ by operator $f$. This revenue formulation assumes if a user traverses two separate links owned by the same operator along a route, they would only pay $p_{rf}$ to that operator once. For a user flow assignment and link operation solution to Eq. (4) there exists one or more corresponding path flow solutions. For a given path flow $\{z_r^*\}_{r \in R}$ we define the outcomes based on this flow, $\big((u,p),z\big)$.

Stability conditions need to ensure that no player in a coalition has incentive to generate a higher payoff by forming another coalition, whether it is a user with another feasible route or an operator with other users in the network. A feasible outcome is presented in **Definition 1**.

**Definition 1.** *The outcome $\big((u,p);z\big)$ is feasible for the corresponding matching problem in Eq. (4) if:*

$$u_s + \sum_{f \in F_r} p_{rf} = U_s - \sum_{(i,j) \in A_r} t_{ij}, \forall\, r \in R_s^*, s \in S \tag{5a}$$

$$u_s \geq 0, \quad \forall\, s \in S \tag{5b}$$

$$p_{rf} \geq 0, \quad \forall\, r \in R, f \in F \tag{5c}$$

$$\sum_{r \in R_f} p_{rf} z_r^* \geq \sum_{(i,j) \in A_f} c_{ij} y_{ij}^*, \quad \forall\, f \in F \tag{5d}$$

where Eq. (5a) ensures that the cost allocations are divided from the surplus of the utility less the travel disutilities and costs (which is shown in Rasulkhani and Chow (2019) to be equivalent to the $a_{ij}$ in Shapley and Shubik (1971). Eq. (5b) and (5c) make sure the cost allocations are non-negative. Eq. (5d) requires the operator payoffs to meet the operating cost threshold, where $y_{ij}^*$ is the optimal binary decisions from the matching problem for an operator to serve a link $(i,j)$ and $z_r^*$ is the flow on an optimal path $r \in R_s^*$. In the case where subsidies $\gamma_{ij} \leq c_{ij}$ exist, Eq. (5d) may have a lower threshold $\big(\sum_{r \in R_f} p_{rf} z_r^* \geq \sum_{(i,j) \in A_f} (c_{ij} - \gamma_{ij}) y_{ij}^*\big)$.

The condition ensuring a stable outcome space that is feasible in Eq. (5a) – (5d) is shown in Eq. (6) for any given path $r' \in R_s \setminus R_s^*$ with respect to each path $r \in R_s^*$.

$$u_s + \sum_{f \in (F_r \cap F_{r'})} p_{rf} \geq U_s - \sum_{(i,j) \in A_{r'}} \left(t_{ij} + \mu_{ij}^* + c_{ij}(1 - y_{ij}^*)\right), \tag{6}$$

$$\forall r' \in R_s \setminus R_s^*, \quad r \in R_s^*, \quad s \in S$$

The stability condition of Eq. (6) ensures that travelers and operators costs are transferred between operators and users such that neither have incentive to deviate from the optimal assignment given by the matching problem (4a) - (4e). The user payoff is denoted by the surplus $u_s$ and $\sum_{f \in (F_r \cap F_{r'})} p_{rf}$ is the profit of operators earned on the optimal path $r \in R_s^*$.

**Proposition 1**. *The stability condition for the matching problem in Eq. (4), given feasibility conditions in Eq. (5), is expressed as Eq. (6).*

***Proof.***

The proof requires demonstrating that Eq. (6) satisfies stability in the same way that the $(u, v; x)$ stability is achieved with $u_i + v_j \geq a_{ij}$ for unmatched pairs in the basic assignment game from Shapley and Shubik (1971).

Consider an unmatched path $r' \in R_s \setminus R_s^*$. As shown in Rasulkhani and Chow (2019), the $a_{ij}$ in the RHS in Shapley and Shubik's model can be represented as the utility less the travel disutility. For an unused path to be switched over to, the disutility needs to consider two other conditions.

*Condition 1*: If a link on the unused path has binding capacity (because of use by another path), the cost to switch over requires moving a passenger off the binding link to another link. This cost is the dual variable $\mu_{ij}$ corresponding to a binding capacity at link $(i, j)$, where it is equal to zero when capacity is nonbinding. The $\mu_{ij}^*$ can be found after solving Eq. (4) by setting the values $y_{ij}^*$ in Eq. (4a) – (4d) constant and finding the capacity dual variables in the LP subproblem.

*Condition 2*: If a link is not matched to any paths (an unmatched path may have links matched to other paths), then the cost of switching to that unmatched path needs to include a fixed charge of $c_{ij}$. The condition of a link $(i, j)$ not being matched to any other path is indicated by the optimal solution $y_{ij}^*$ from the matching problem in Eq. (4). As such, the RHS surplus utility switching to an unmatched path is shown in Eq. (7).

$$U_s - \sum_{(i,j) \in A_{r'}} \left( t_{ij} + \mu_{ij}^* + c_{ij}(1 - y_{ij}^*) \right), s \in S, r' \in R_s \setminus R_s^* \tag{7}$$

The payoffs in the LHS $(u_i + v_j)$ of the stability condition represent what the operators and user group get when matched. For user group $s$, this is simply $u_s$. The payoffs for operators $f \in R_s \setminus R_s^*$, however, need to consider the fares that they currently get. The payoff should be the supremum of all the payoffs $\sup_{r \in R_s^*} \delta_{rf} p_{rf}$, where $\delta_{rf} = 1$ if operator $f$ is on path $r$, the minimum amount needed to switch one user from one of the optimal paths operated by $f$ to the new path. When evaluating all operators and adding the user payoff, this becomes Eq. (8).

$$u_s + \sum_{f \in F_{r'}} \sup_{r \in R_s^*} \delta_{rf} p_{rf} \tag{8}$$

Combining with Eq. (7) leads to the following inequality in Eq. (9).

$$u_s + \sum_{f \in F_{r'}} \sup_{r \in R_s^*} \delta_{rf} p_{rf} \geq U_s - \sum_{(i,j) \in A_{r'}} \left(t_{ij} + \mu_{ij}^* + c_{ij}(1 - y_{ij}^*)\right), s \in S, r' \in R_s \setminus R_s^* \quad (9)$$

The supremum is nonlinear but can be removed by replacing Eq. (9) with a set of constraints comparing $r' \in R_s \setminus R_s^*$ with each $r \in R_s^*$. In addition, the following relationship for the set of prices holds in Eq. (10).

$$\sum_{f \in F_{r'}} p_{rf} = \sum_{f \in F_{r'} \cap F_r} p_{rf} + \sum_{f \in F_{r'} \setminus F_r} p_{rf} \quad (10)$$

$\sum_{f \in F_{r'} \setminus F_r} p_{rf} = 0$ since these prices belong on the unmatched paths. Eq. (9) is then equivalent to the linear set of Eq. (6). ∎

The stable outcome problem is then defined as a mathematical program with linear constraints formed by Eq. (5) – (6). The objective can be set to maximize user cost allocation (seller-optimal: $\max Z = \sum_f \sum_{r \in R} p_{rf} z_r$) or to maximize operator cost allocation (buyer-optimal: $\max Z = \sum_s u_s$), both of which lead to linear objectives. Since the first two objectives are linear, determining the stable outcome space between them can therefore be done using linear programming.

### *3.1.3. Model properties and limitation considerations*
The proposed model is static and designed for planning purposes. Much like how dynamic traffic can be modeled by static models for some purposes (e.g. link expansions and pricing decisions) but dynamic models are needed for other purposes (e.g. dynamic route guidance technologies or traffic control strategies), this model can be used to capture certain aspects of dynamic MaaS services like ride-hail, demand responsive transit, or shared vehicles. For example, shared vehicles fundamentally operate in a dynamic manner with capacities that change throughout the day based on pickups and drop-offs and rebalancing strategy. Nonetheless, one can observe such a system over multiple days and use the maximum flows between nodes to model the steady state of such a system as a complete graph with those steady state capacities. Such a model can be used to evaluate the propensity for the service to serve traveler demand, although it would not be adequately sensitive to comparing rebalancing strategies in an operational setting. Like the progression of traffic assignment toward dynamic traffic assignment, a dynamic assignment game model would be needed for this purpose.

Proper calibration of the model parameters is essential to provide empirical validation. The link travel costs should correspond to average undersaturated flow travel costs. Capacity may be obtained from service frequency or based on observed steady state maximum flows. Operating costs can also be obtained from data directly. The most challenging parameter to calibrate is the utility $U_s$, which is generally a latent and heterogeneous attribute. One way to calibrate it is to use an existing market scenario. That was the approach adopted by Rasulkhani and Chow (2019) in their case study of NYC taxis, using the existing operations to determine the lower bound utilities.

A better approach is to estimate the utilities with a random utility model estimated from data. Utility can be modeled as the sum of the generalized route cost, a constant term (which is also calibrated) plus a random error term: $U_{sr} = U_s^0 + t_{sr} + \varepsilon_{sr}$. We investigated this in a separate study of applying the model from Rasulkhani and Chow (2019) to real data in Ma et al. (2020). In that study, stochastic utility was used to account for heterogeneous user preferences in an assignment game. The challenge in that approach is that the stability condition previously used is no longer applicable, and alternative stochastic stability conditions are needed (e.g. $\epsilon$-stability (Wang et al., 2018)). We intend to investigate such extensions for the proposed model in future research.

## 3.2. Closed form examples using proposed model

The model is used to provide a closed-form solution to stylized settings to illustrate its capabilities. Two cases are shown in **Figure 3**.

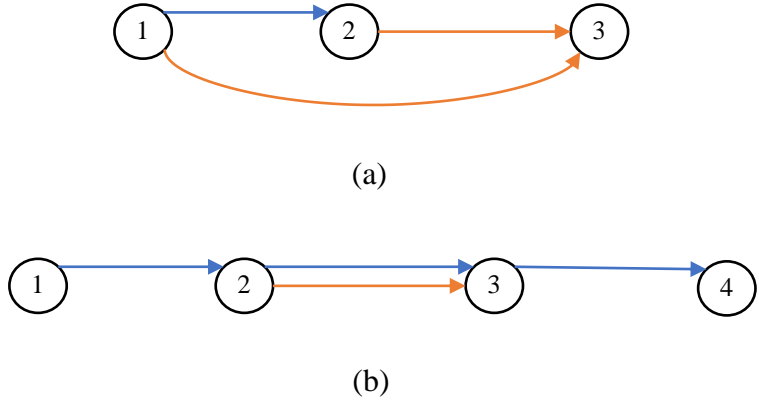

(a)

(b)

**Figure 3**. (a) choosing between cooperation and competition; (b) small operator against a larger operator.

### 3.2.1. Case 1: Choosing cooperation over competition

In this example, the orange operator can choose between serving a link in competition or one in cooperation. In order to serve users travelling from node 1 to node 3 the orange operator must choose between operating link 13 directly or cooperating with the blue operator and operate link 23. Suppose capacity is sufficiently large for all links.

**Lemma 1**. *In a 2-operator game as shown in **Figure 3a**, let $y_{12}$ be the binary decision for an operator without the option of providing solo service; and let $y_{23}$ be the cooperation decision and $y_{13}$ be the solo operating decision of a separate operator. A lower bound for the stable outcome for the latter operator to enter a cooperative multimodal solution with the former operator ($y_{12} = y_{23} = 1$) instead of providing solo service ($y_{13} = 1$) is shown in Eq. (11).*

$$p_o \geq \frac{c_{12} + c_{23}}{d} - t_{13} - c_{13} + t_{12} + t_{23}, \qquad \text{if } t_{12}d + t_{23}d + c_{12} + c_{23} \leq t_{13}d + c_{13} \quad (11)$$

The proof is Appendix A.

**Lemma 1** provides a bound that is independent of the cost allocation of the other operator or users, or their utilities. It provides insight for operators that have a choice of operating alone or teaming with other operators that may not have the option to choose to serve solo. This may prove insightful when considering MaaS markets and when it is worthwhile for an operator to team with another, particularly in fulfilling last mile service.

### 3.2.2. *Case 2: Bargaining power of small operators (orange) serving a segment of a larger transit line (blue)*

Consider **Figure 3b**, which represents a smaller operator (orange) serving one segment of a multi-segment route operated by a larger operator (blue). There may be multiple OD pairs, but given that there's only one segment that differs, we can express all the OD pairs equivalently as one OD pair for (1,4) with two paths. Path 1 is operated by the blue operator without making any transfers. Path 2 involves having travelers transfer at node 2 to the orange operator and back to the blue at node 3.

We study the pricing power of a smaller operator (orange) compared to a larger (blue) that has more access to right of way. We assume that capacity is not binding, i.e. $\mu_{23}^{blue} = 0$. Path 2 is optimal when: $t_{23}^o \leq t_{23}^b$, where $t_{23}^o$ (travel time for orange operator) may include transfer times.

**Lemma 2**. *For the 2-operator assignment game in **Figure 3b** with demand for (1,4) and non-binding capacities, the upper bound on the stable outcome for a smaller operator is shown in Eq. (12) if $t_{23}^o \leq t_{23}^b$.*

$$p_{2o} \leq \begin{cases} t_{23}^b - t_{23}^o + c_{23}^b, & \text{if } x_{23}^b = 0 \\ 0, & \text{if } x_{23}^b > 0 \end{cases} \tag{12}$$

The proof of **Lemma 2** is Appendix B. The lemma suggests that a small operator can only reach a positive stable outcome if they are strictly improved in travel time than the larger operator, and the pricing they set cannot exceed the difference in performance. For example, as the travel time difference $t_{23}^b - t_{23}^o$ decreases toward zero, the upper bound is primarily influenced by the cost of the larger operator. Noting the feasibility condition $p_{2o}d > c_{23}^o$, then a stable outcome can only exist for this extreme setting if $\frac{c_{23}^o}{d} < c_{23}^b$. On the other hand, if the links are collectively so long that there is room for $t_{23}^b - t_{23}^o$ to dominate, then a stable outcome can exist with $\frac{c_{23}^o}{d} < t_{23}^b - t_{23}^o$.

### 3.3. Non-uniqueness discussion

Since we propose to use this model as a descriptive tool for policymakers, the issue of non-uniqueness needs to be discussed. An outcome depends on a path flow assignment. However, a path flow solution is non-unique to Eq. (4). Due to the non-uniqueness of paths in the matching solution, the set of operator-routes that enter the market at equilibrium, and the corresponding cost allocations, is not unique. As discussed in Rasulkhani and Chow (2019), many-to-many assignment games (of which this model belongs) are known not to have one-to-one correspondence between the matching solutions and the outcomes. Still, some quantities (and their associated measures) are unique: e.g. link flows and sum of capacity dual variables along each path, and consequently the total revenues that can be gained by all operators and total consumer surplus gained by users are also unique. These measures are still useful for comparing between alternative platform designs.

In addition, the outcomes depend on the selection of links to serve (the $y_{ij}^*$ in Eq. (6)), the path flows ($z_r^*$ in Eq. (5d)), and the capacity dual variables (the $\mu_{ij}^*$ in Eq. (6)). This means that there is a unique stable outcome space for a given matching, even if we cannot guarantee the opposite. **Figure 4** illustrates how two different matches may lead to one outcome space each. For example, we might assume a market is under Match 1 and quantify its measures based on Outcome Space 1. If such an outcome is implemented based on this result but it turns out to be stable for another optimal Match 2, the solution is still optimal, but the measures may not reflect values from original assumed Match 1. However, Match 2 should still lead back to Outcome Space 1.

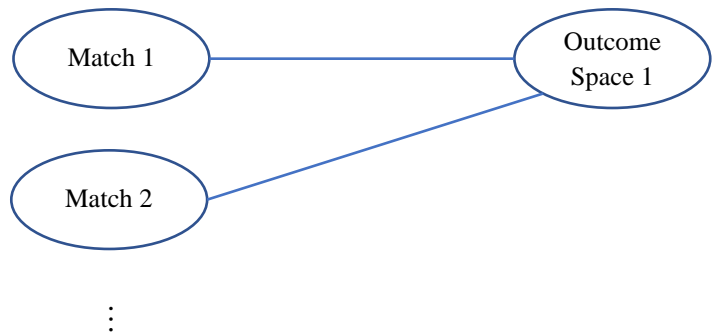

**Figure 4**. The many-to-one association of matches to outcomes due to the non-unique matches in the many-to-many assignment game.

The implication of this many-to-one correspondence between optimal matches and outcome spaces is that sampling of MaaS operators' user data to obtain the realized MaaS paths after implementation (which is possible under a unified platform that tracks service usage through mobile apps, for example) provides a means to update the measures to the new matching solution. In this latter use, the model provides *ex post* evaluation.

In summary, many-to-many assignment games like the proposed model are limited in the interpretation of stable outcomes because of non-uniqueness. Nonetheless we can still learn much in terms of unique link measures, aggregate measures, and can justify using the model in ex post analysis to explain effects.

## 4. CONSTRAINT GENERATION ALGORITHM USING LEXICOGRAHPIC CORE ALLOCATIONS

It is not practical to explicitly enumerate all feasible paths $R_s$ to construct the stability constraints in Eq. (6). We develop a method similar to Bahel and Trudeau (2014) to identify only the paths that form the extreme boundaries of the shortest path problem cooperative game to determine a subset $\tilde{R}_s \subset R_s$ that can produce an equivalent set of constraints to Eq. (6). The algorithm leads to exact solutions.

### 4.1. Solution algorithm

We propose an algorithm that generates a subset of feasible routes. This subset is chosen such that they form the extreme points of a core of a shortest path game. A shortest path game involves a set of operators forming coalitions to provide shortest paths to connect sources to sinks. The core is the set of cost allocations that ensure no player in a coalition would break away. Fragnelli et al. (2000) provide an overview of this class of problems.

In generating the alternative paths, we define a cost for a set of links corresponding to a shortest path game as $\omega_r$ shown in Eq. (13).

$$\omega_r = \sum_{(i,j) \in A_r} \left( t_{ij} + \mu_{ij}^* + c_{ij}(1 - y_{ij}^*) \right), \forall r \in R \tag{13}$$

We express a solution to the shortest path problem (SPP) in coalitional form. The optimal path that corresponds to a user $s \in S$ is $r = \arg\min\{\omega^s(r') | r' \in R_s\}$. The coalition formed is $V^* = F_r \cup \{s\}$. Given an optimal path $r \in R_s^*$ and the set of operators $F_r$, the extreme points of the core are defined by $a^\pi$ for a lexicographically ordered subcoalition $\pi \in \Pi(F_{R_s^*})$. These permutations are expressed by a set of indexed operators that are listed in lexicographical ordering shown in Eq. (14).

$$\emptyset \neq \{f \in F_r\} = \{f_1, \dots, f_l\} \tag{14}$$

where $l \geq 1$ and $\pi(f_1) < \cdots < \pi(f_l)$. For example, given an optimal path r that corresponds to a set of three operators $F_r = \{1,2,3\}$ such that the ordering of operators along the path is [2,1,3]. Regardless of the ordering, the corresponding lexicographically ordered subcoalition is $\pi = [1,2,3]$ with $\pi_{f_1} = 1$.

Next, we propose a decomposition technique similar to Bahel and Trudeau (2014) to derive the allocations that lie at the extreme points of the core $a^\pi$ associated with $\pi \in \Pi(F_{R_s^*})$ that correspond to a set of optimal paths $R_s^* \forall s \in S$. Our algorithm differs from Bahel and Trudeau (2014) in two ways. They consider network nodes to be owned by a set of agents and focuses on logistics networks and use the algorithm to produce stable shortest paths. However, in our study we consider multimodal capacitated links. The paths generated are used as the equivalent replacement of a fully enumerated path set for determining the stable outcome subproblem constraints.

**Algorithm 1** generates stability constraints for each user in the network (**step 1**). If the set $R_s^*$ contains more than one path, then: $F_{R_s^*} = \cup F_r \, \forall r \in R_s^*$ (**step 2 & 3**). The algorithm generates optimal paths (**step 5**) without the participation of any subset (permutation) of operators found in the set of optimal paths $R_s^*$ (**step 4**).

**Algorithm 1. Constraint generation for Eq. (6) without explicit path enumeration**

1. **For** each user in the set $S$ **do**
2.     Set $R_s^* = \{r \in R_s | \omega(r) \leq \omega(r') \; \forall r' \in R_s\}$
3.     **For** each optimal path $r \in R_s^*$ **do**
4.         **For** $\pi$ in $\Pi(F_{R_s^*})$ **do**
5.             Generate $u_s + \sum_{f \in (F_r \cap F_{\bar{r}_\pi})} p_{rf} \geq U_s - \omega^s(\bar{r}_\pi)$,
                  where $\bar{r}_\pi = \arg\min\{\omega^s(r) | r \in R_s \; s.t. \; \pi \cap F_{R_s^*} = \emptyset\}$

In identifying the feasible paths $R_s$ for creating the stability constraints Eq. (6), a path in which the cost allocations lie outside of the core need not be compared to if there's a better path to compare against, as presented in **Proposition 2**.

**Proposition 2**. *The generation of stability constraints for each optimal path $r \in R_s^*$ using **Algorithm 1** is equivalent to the constraints in Eq. (6) formed from $R_s$.*

**Proof**. An allocation for the shortest path problem $R_s = (F, \omega, d_s)$ and a permutation $\pi$ of an optimal path is given by $a^\pi(R_s)$. To establish that the constraint generation algorithm is equivalent, we need to show that $a^\pi(R_s)$ is an extreme point of the core of the SPP. The allocation $a^\pi$ is well defined since: $a_s^\pi + \sum_{i=1}^{l} a_{f_i}^\pi = U_s - \omega_r$ and $a_f = 0$ if $f \notin \{f_1, \ldots, f_l\}$. First, $a^\pi$ is shown to be stable when no sub-coalition $V \subseteq V^*$ can achieve a higher profit in another coalition than through their share in $a^\pi(V)$.

    1) If the traveler $s \notin V$ then obviously no coalition can improve on $a^\pi$, since $a^\pi(V) \leq 0$.

    2) If a path does not include any of the operators that lie in the optimal path set: $s \in V$ and $V \setminus \{s\} \subseteq F \setminus \{f_1, \ldots, f_l\}$ then:

$$a^\pi(V) = a_s^\pi = (U_s - \omega(r)) - a_{f_1}^\pi - \cdots - a_{f_l}^\pi = U_s - \omega^s(\bar{r}_\pi) \geq U_s - \omega(r') \; \forall r' \in R_s \; s.t. \; F_{r'} \subseteq F \setminus \{f_1, \ldots, f_l\}.$$

    3) For the case that a path includes any subset of operators that also belongs to the optimal path operator set: $s \in V$ and $(V \setminus \{s\}) \cap \{f_1, \ldots, f_l\} = \{k_1, \ldots, k_t\}$, $1 \leq t \leq l$ and $\pi(k_1) < \cdots < \pi(k_t) < \pi(s)$ we get:

$$a^\pi(V) = a_{k_1}^\pi + \cdots + a_{k_t}^\pi + a_s^\pi = a_{k_1}^\pi + \cdots + a_{k_t}^\pi + (U_s - \omega^s(\bar{r}_\pi))$$

Assume by contradiction that there exists a higher payoff path $r' \in R_s \; s.t. \; F_{r'} \subseteq V$:

$$U_s - \omega_{r'} > a^\pi(V) = a_{k_1}^\pi + \cdots + a_{k_t}^\pi + (U_s - \omega^s(\bar{r}_\pi))$$

The number of permutations is $n_{t'} \equiv |\{f \in F_{r'} \; s.t. \; \pi(f) < \pi(k_{t'})\}| + 1$ for $1 \leq t' \leq t$. Then we can write that:

$$U_s - \omega_{n_1}(r': f_{k_1} \notin F_{r\prime}) \equiv (U_s - \omega_{n_1-1}(r')) - a_{k_1}^{\pi} \text{ and since } \omega_{n_1-1}(r') \leq \omega(r'):$$

$$\omega_{n_1-1}(r') - a_{k_1}^{\pi} \leq \omega(r') - a_{k_1}^{\pi} < a_{k_2}^{\pi} + \cdots + a_{k_t}^{\pi} + \omega^s(\bar{r}_\pi)$$

For $k_2$, we repeat the same process:

$$\omega_{n_2}(r': f_{k_1}, f_{k_2} \notin F_{r\prime}) \leq \omega_{n_1}(r': f_{k_1} \notin F_{r\prime}) - a_{k_2}^{\pi} < a_3^{\pi} + \cdots + a_{k_t}^{\pi} + \omega^s(\bar{r}_\pi)$$

For the $t^{th}$ repetition, we get Eq. (15).

$$\omega_{n_t}(r': f_{k_1}, \ldots, f_{k_t} \notin F_{r\prime}) \leq \omega_{n_{t-1}}(r': f_{k_1}, \ldots, f_{k_{t-1}} \notin F_{r\prime}) - a_{k_t}^{\pi} < \omega^s(\bar{r}_\pi) \quad (15)$$

Given that $f_q \notin V$, for any $f_q \in F_r$ s.t. $\pi(f_q) > \pi(k_t)$ we can conclude that:
$\omega_q(r': f_{k_1}, \ldots, f_{k_t} \notin F_{r\prime}) = \omega_{n_t}(r': f_{k_1}, \ldots, f_{k_t} \notin F_{r\prime})$, for any $q = n_t, \ldots, l$ and
$\omega_l(r': f_{k_1}, \ldots, f_{k_t} \notin F_{r\prime}) = \omega_{n_t}(r': f_{k_1}, \ldots, f_{k_t} \notin F_{r\prime})$. From Eq. (15) we get:

$$\omega_l(r': f_{k_1}, \ldots, f_{k_t} \notin F_{r\prime}) < \omega^s(\bar{r}_\pi)$$

This is a contradiction since $\omega^s(\bar{r}_\pi)$ is a minimal cost path. This concludes the proof that the constraint bound $\omega(\bar{r}_\pi)$ generated by: $\bar{r}_\pi = argmin\{\omega(r) | r \in R_s \text{ s.t. } \pi \cap F_r = \emptyset\}$ corresponds to permutation the extreme point of the core $a^{\pi}$, which implies that $\bar{\omega}_l^s(\pi)$ is the minimal cost corresponding to the core vertices which includes operators $f \in \pi$. This means finding only constraints corresponding to the extreme points is equivalent to constraints formed from explicit enumeration of all candidate paths. ∎

Bahel and Trudeau's (2014) proof was applied to generate core extrema. Our proof explicitly deals with the equivalency of the constraint sets. **Algorithm 2** provides a summary of the entire solution procedure that was described in Sections 3 and 4.

**Algorithm 2. Solution method for finding stable outcomes of many-to-many assignment game**

1. Solve the matching problem as an MCND using an existing solution algorithm, e.g. path-based column generation algorithm (Gendron and Larose, 2014), to obtain links operated $y_{ij}^*$ and a set of user paths $R_s^*$.
2. Fix the values of $y_{ij}^*$ and solve the LP subproblem of Eq. (4) to obtain link capacity dual variables $\mu_{ij}^*$ and path flows $z_r^*$.
3. Construct the stable outcome problem depending on objective function. The constraints of the program include the linear feasibility conditions in Eq. (5) and the linear stability conditions in Eq. (6). The extreme points of the core of $R_s$ are used to generate the stability constraints (Eq. (6)) using **Algorithm 1**.

4. Solve the stable outcome problem for the buyer-optimal and seller-optimal extremes to determine the range of stable outcomes corresponding to the matching solution.

## 4.2. Illustrative instance

An example of the model is shown in **Figure 5** for a market with node 2 decomposed into transfer links ((21,22), (21,23)) not owned by any operator. There are two OD pairs, each with utilities of $U_s = 20$ and demand $d_{13} = 1000$ and $d_{14} = 500$. Operator A (blue) owns links {(1,3), (1,21)}, operator B (orange) owns (22,3), operator C (green) owns (23,4), operator D (red) owns (1,4), operator E (purple) has (1,5) and (5,4), and operator F (yellow) (1,6) and (6,4). The optimal assignment from the matching problem has 1000 flow on (1,3), 200 flow on (1,21,23,4), and 300 flow on (1,4). The grey colored links represent transfers between operators. The capacity at link (1,21) is binding with a corresponding dual variable $\mu_{12} = 4$.

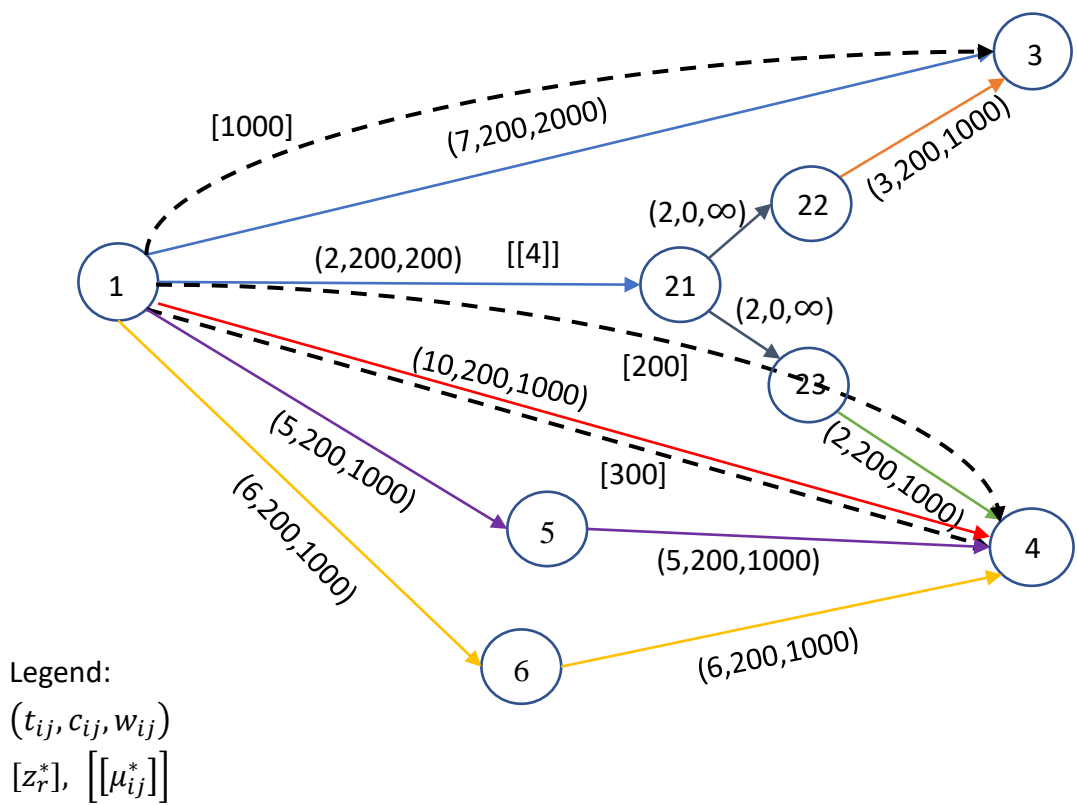

Legend:
$(t_{ij}, c_{ij}, w_{ij})$
$[z_r^*], [[\mu_{ij}^*]]$

**Figure 5**. Example network with 6 nodes, 6 operators, and 2 OD pairs assigned to 6 paths.

**Algorithm 2** is illustrated to generate stability constraints for the stable outcome problem. The path flows are $z_1 = 1000, z_2 = 200, z_3 = 300$ corresponding to the dashed lines in **Figure 5**, where path 1 is (1,3), path 2 is (1,21,23,4), and path 3 is (1,4). The operator set is $F = \{A, B, C, D\}$. Constraints (5a) – (5d) for this example are reflected as Eq. (16) – (23).

$$p_{1,A} + u_{(1,3)} = 13 \quad (16)$$

$$p_{2,A} + p_{2,C} + u_{(1,4)} = 14 \tag{17}$$
$$p_{3,D} + u_{(1,4)} = 10 \tag{18}$$
$$1000 p_{1,A} + 200 p_{2,A} \geq 400 \tag{19}$$
$$200 p_{2,C} \geq 200 \tag{20}$$
$$300 p_{3,D} \geq 200 \tag{21}$$
$$p_{rf}, u_s \geq 0 \tag{22}$$
$$p_{r0} = 0 \tag{23}$$

The steps of **Algorithm 1** are then shown below.

1. For the user $(1,3)$ we have only $r = (1,3)$ as the optimal path and $F_r = \{A\}$
    - The set of permutations: $\pi = \{A\}$
    - We solve the problem: $\bar{r}_\pi = argmin\{\omega^s(r) | r \in R_s \ s.t. \ \pi \cap F_r = \emptyset\}$ and obtain $\bar{r}_\pi = \emptyset$, since the alternative path of link (1,21) also belongs to operator A.

2. For the user $(1,4)$ we have $R_{(1,4)} = \{(1,23,24,4), (1,4)\}$ as the optimal path set.
    - The operator set includes operators from both paths $F_{R_{(1,4)}} = \{A, C, D\}$ (we do not include the transfer link).
    - $\Pi(F_{R_{(1,4)}}) = \{(A), (C), (D), (A,C), (A,D), (C,D), (A,C,D)\}$.
    - For cases when $\pi: \pi \setminus \{(A,C)\}$, $\bar{r}_\pi = (1,23,24,4)$ and $\pi: \pi \setminus \{(D)\}$, $\bar{r}_\pi = (1,4)$, we do not need to generate stability constraints for optimal paths since Eqs. (17) – (18) require equal user payoff for both routes $u_{(1,4)}$.
    - For the cases: $\pi \in \{(A,C,D), (A,D), (C,D)\}$, $\bar{r}_\pi = (1,5,4)$ and $\omega(\bar{r}_\pi) = 7 + 3 + \underbrace{200}_{c_{15}} + \underbrace{200}_{c_{54}}$ and optimal path $r = (1,23,24)$ we generate Eq. (24) as an equivalent to Eq. (6).

$$u_{(1,4)} \geq -390 \tag{24}$$

The algorithm avoids including the stability constraint for the alternative feasible path (1,6,4), $u_{(1,4)} \geq -392$, since that is not part of the extrema. That constraint would be dominated by Eq. (24).

Based on the constraints, the buyer-optimal and seller-optimal allocations are shown in **Table 2**. Operator B does not enter the market. Since there are no identical non-binding, used paths in this example, the path flow solution is unique, and we can therefore get unique revenues per operator. For this mix of operated links, Operator A stands to gain net profits between 333.33 to 15,600, Operator C can only gain up to 200 and Operator D up to 3000. As seen from these results, Operator A has the most profit to gain due to negotiating power from the binding capacity in link (1,21). At this stage, the operators and users (through the city agency as proxy) can work out the cost allocation mechanism that falls within the convex outcome range.

From this example, we can extract sensitivity of each operator's performance and consumer surplus based on changes in link capacities, OD demand, utility per OD pair (which relates to users' preference relative to other mobility options outside of the platform), operating cost, generalized travel disutility (alterations in-vehicle time, access time, transfer time, wait time due to external factors or operator policies), and addition/removal of operator candidate links.

**Table 2**. Solution to example M2M problem

| (Operator,user-route) | Flow $\sum_{r \in R_f} z_r$ | Buyer-optimal Price $p_{rf}$ | Seller-optimal Price $p_{rf}$ |
|---|---|---|---|
| (A, (1,3)) | 1000 | $0 | $13 |
| (A, (1,21,23,4)) | 200 | $3.67 | $13 |
| (C, (1,21,23,4)) | 200 | $1.0 | $1 |
| (D, (1,4)) | 300 | $0.67 | $10 |
| **User group-route** | $U_s - \sum_{(i,j) \in A_r} (t_{ij})$ | $U_s - \sum_{(i,j) \in A_r} (t_{ij}) - \sum_{f \in F_r} p_{rf}$ | $U_s - \sum_{(i,j) \in A_r} (t_{ij}) - \sum_{f \in F_r} p_{rf}$ |
| ((1,3),(1,3)) | $13 | $13 | $0 |
| ((1,4),(1,21,23,4)) | $14 | $9.33 | $0 |
| ((1,4),(1,4)) | $10 | $9.33 | $0 |
| **Operator** | | $\sum_{r \in R_f} p_{rf} z_r - \sum_{(i,j) \in L_f} C_{ij} y_{ij}$ | $\sum_{r \in R_f} p_{rf} z_r - \sum_{(i,j) \in L_f} C_{ij} y_{ij}$ |
| A | | $333.33 | $15,600 |
| B | | $0 | $0 |
| C | | $0 | $200 |
| D | | $0 | $3000 |
| E | | $0 | $0 |
| F | | $0 | $0 |

## 5. NUMERICAL TESTS

Having illustrated the model and algorithm, we test the effectiveness of the method on a larger instance, the classic 24-node Sioux Falls network (see Stabler, 2019) as shown in **Figure 6a**. In this duopoly there are two operators: Operator 1 (blue) is a bus network while Operator 2 (orange) is a set of two rail lines. Transfer links are denoted by grey links (28 additional links added, 14 nodes added, for total of 38 nodes and 104 links). Since a network of even this size is too large for explicit path enumeration for the 552 OD pairs, we divide the tests into two sections:
- Section 5.1 takes a subnetwork of Sioux Falls with only 4 OD pairs that can be solved using the explicit path enumeration to compare to the effectiveness of the proposed algorithm.
- Section 5.2 takes the full Sioux Falls network with added transfer links to test the scalability of the algorithm.

### 5.1. Sioux Falls subnetwork

#### *5.1.1. Experimental design*
We consider two operators in the base scenario. For this base scenario, we assume that both operators have revenue maximization ($\max Z = \sum_{r \in R_f} p_{rf} z_r$) as their objective. Relative to this baseline, we seek several experimental objectives:

1) Assessing the new prevailing market conditions and payoff allocations when one of the two operators is acquired by a **government agency** and becomes welfare maximizing.
2) Evaluate the **entry of a new operator** to the consumer surplus and market revenues of existing operators.
3) Evaluate the impact of one operator increasing its **binding capacity** on the consumer surplus and market revenues of other operators.
4) Evaluate the effect of **technology improvement** (e.g. improved matching algorithms) that reduce operating costs $c_{ij}$ for privately operated links while increasing travel times $t_{ij}$ for the whole network.

The free flow travel times from Sioux Falls are used as the $t_{ij}$ and set the same as $c_{ij}$. The cost units are all in $. The parameters are listed in **Table 3**.

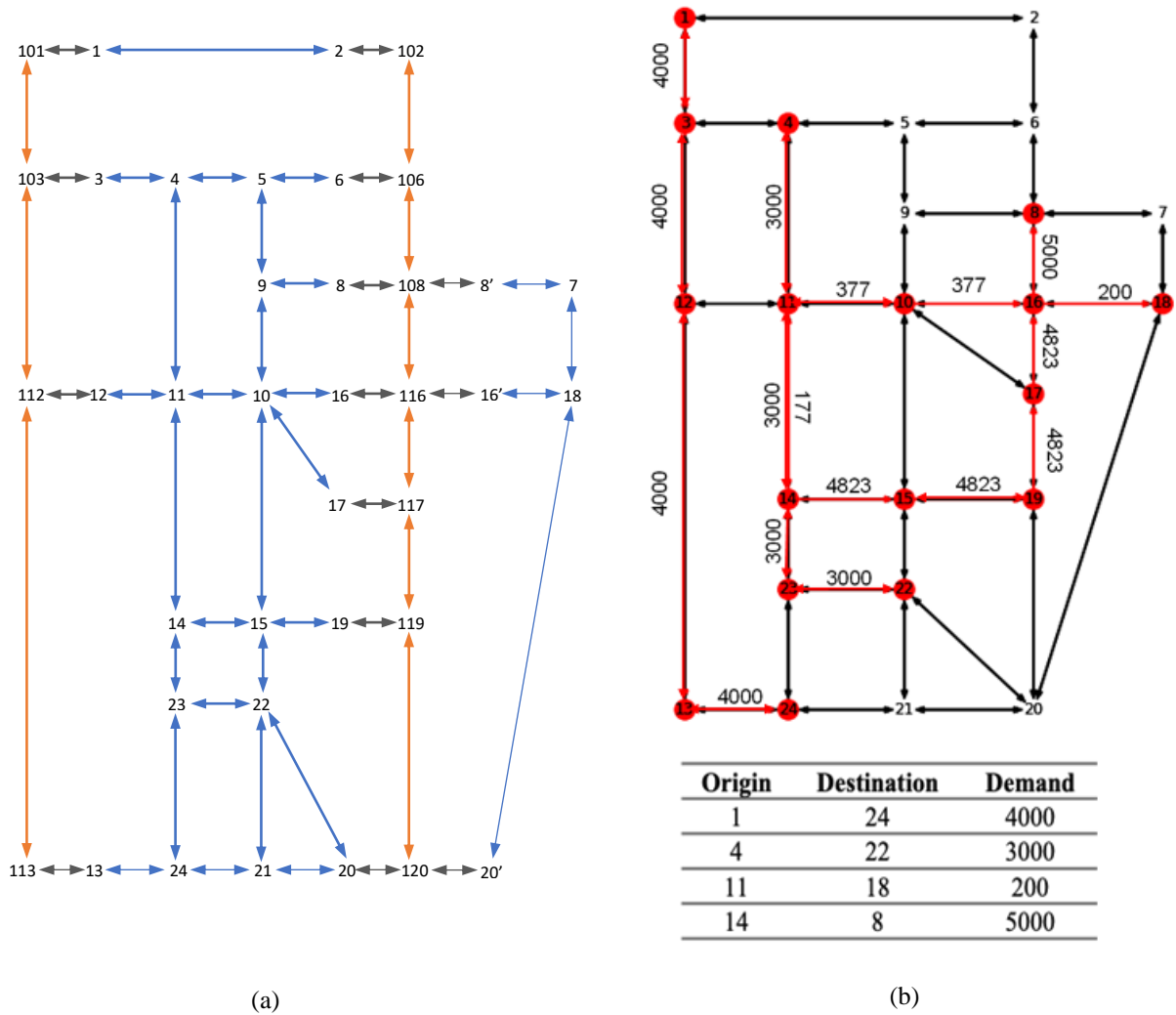

(a)   (b)

**Figure 6**. Sioux Falls (a) network with transfer links (grey), a rail operator (orange), and a bus transit operator (blue); and (b) Eq. (4) assignment under modified subnetwork OD demand with 0 cost transfers.

**Table 3.** Parameters for Sioux Falls examples

| (i,j) | $w_{ij}$ | $c_{ij}/t_{ij}$ | (i,j) | $w_{ij}$ | $c_{ij}/t_{ij}$ | (i,j) | $w_{ij}$ | $c_{ij}/t_{ij}$ | (i,j) | $w_{ij}$ | $c_{ij}/t_{ij}$ |
|---|---|---|---|---|---|---|---|---|---|---|---|
| (1,2) | 25901 | 6 | (8,7) | 7842 | 3 | (13,24) | 5092 | 4 | (119,117) | 4824 | 2 |
| (101,103) | 23404 | 4 | (8,9) | 5051 | 10 | (14,11) | 4877 | 4 | (119,120) | 5003 | 4 |
| (2,1) | 25901 | 6 | (108,116) | 5046 | 5 | (14,15) | 5128 | 5 | (20,18) | 23404 | 4 |
| (102,106) | 4959 | 5 | (9,5) | 10000 | 5 | (14,23) | 4925 | 4 | (120,119) | 5003 | 4 |
| (103,101) | 23404 | 4 | (9,8) | 5051 | 10 | (15,10) | 13513 | 6 | (20,21) | 5060 | 6 |
| (3,4) | 17111 | 4 | (9,10) | 13916 | 3 | (15,14) | 5128 | 5 | (20,22) | 5076 | 5 |
| (103,112) | 23404 | 4 | (10,9) | 13916 | 3 | (15,19) | 14565 | 3 | (21,20) | 5060 | 6 |
| (4,3) | 17111 | 4 | (10,11) | 10000 | 5 | (15,22) | 9600 | 3 | (21,22) | 5230 | 2 |
| (4,5) | 17783 | 2 | (10,15) | 13513 | 6 | (116,108) | 5046 | 5 | (21,24) | 4886 | 3 |
| (4,11) | 4909 | 6 | (10,16) | 4855 | 4 | (16,10) | 4855 | 4 | (22,15) | 9600 | 3 |
| (5,4) | 17783 | 2 | (10,17) | 4994 | 8 | (116,117) | 5230 | 2 | (22,20) | 5076 | 5 |
| (5,6) | 4948 | 4 | (11,4) | 4909 | 6 | (16,18) | 19680 | 3 | (22,21) | 5230 | 2 |
| (5,9) | 10000 | 5 | (11,10) | 10000 | 5 | (17,10) | 4994 | 8 | (22,23) | 5000 | 4 |
| (106,102) | 4959 | 5 | (11,12) | 4909 | 6 | (117,116) | 5230 | 2 | (23,14) | 4925 | 4 |
| (6,5) | 4948 | 4 | (11,14) | 4877 | 4 | (117,119) | 4824 | 2 | (23,22) | 5000 | 4 |
| (106,108) | 4899 | 2 | (112,103) | 23404 | 4 | (18,7) | 23404 | 2 | (23,24) | 5079 | 2 |
| (7,8) | 7842 | 3 | (12,11) | 4909 | 6 | (18,16) | 19680 | 3 | (24,13) | 5092 | 4 |
| (7,18) | 23404 | 2 | (112,113) | 25901 | 3 | (18,20) | 23404 | 4 | (24,21) | 4886 | 3 |
| (108,106) | 4899 | 2 | (113,112) | 25901 | 3 | (19,15) | 14565 | 3 | (24,23) | 5079 | 2 |

### 5.1.2. Computational test for Algorithm 2

The solution to Eq. (4) for the network with $0 cost transfer links is shown in **Figure 6b**, where the red links are the ones that need to be operated. From this solution, only link 58 (node 119 to 117) is at binding capacity resulting in a dual variable of $\mu^*_{58} = 1$. **Table 4** summarizes all these results which are discussed subsequently in each section. Columns with brackets [ ] refer to measures for each operator (blue, orange, and green for the scenario with a new operator).

The last two columns refer to the run times via explicit path enumeration versus the **Algorithm 2** approach. Two numbers are included in each cell; the first number is the time it takes to construct the constraints for the stable outcome model while the latter is the solution time (in milliseconds).

Two significant results are achieved. First, the tests verify that the proposed Algorithm 2 obtains the same results as the path enumeration for Eq. (6). Second, the computational savings when using the constraint generation method instead of path enumeration are highly significant. The problem's solution time is reduced on average by 98%.

### 5.1.3. Rail-bus duopoly

The network duopoly shown in **Figure 6a** is treated as the "base scenario". We assume that the rail service sets a unique cash price for all O-D routes that use it. This "cash fare" policy results in a more constrained price setting than the stability conditions in Eq. (5) – (6) where pricing could vary by user route.

**Table 4.** Comparison of aggregate measures of different scenarios with $U_S = 20$

| Parameters:<br><br>Scenarios: | Revenues ($)<br>[$f = 1, 2, 3$] | Avg. operator fare ($)<br>[$f = 1, 2, 3$] | Avg. operated link revenue ($) | Operator ridership $\sum_{r \in R_f} z_r$<br>[$f = 1, 2, 3$] | Runtime: Model generation / Solution (msec) (original LP) | Runtime: Model generation / Solution (msec) (constraint generation) |
|---|---|---|---|---|---|---|
| **Network duopoly (Base scenario)** | [24424,18000] | [2,2] | 2497 | [12200,9000] | 6259.2 / 20.4 | 28.8 / 0.3 |
| **Government rail acquisition** | [42417,7] | [3.47,0.0008] | 2497 | [12200,9000] | 6259.2 /20.4 | 28.8 / 0.3 |
| **Firm entry** | [60422,10000,0.75] | [5,2,0.0002] | 3912 | [12200,5000,4000] | 12095.6 /35.6 | 32.8 / 0.2 |
| **Binding capacity increase ($w_{58} = 4900$)** | [24500,18000] | [2,2] | 2497 | [12200,9000] | 6259.2 / 20.4 | 28.8 / 0.3 |
| **Binding capacity increase ($w_{58} = 5000$)** | [15600,27000] | [1.3,3] | 2663 | [12200,9000] | 5300.8 / 15.5 | 27.8 / 0.2 |
| **Technological change** | [27506,40500] | [2.25,4.5] | 3400 | [12200,9000] | 6259.2 / 20.4 | 28.8 / 0.3 |

The cost allocation mechanism assumes that both operators seek to maximize their revenues. This translates to an objective value of: $Z = \sum_{r \in R_1} p_{r1} z_r$. The resulting revenue allocation for this example is $R_1 = \$24,424$ for the private mobility service and $R_2 = \$18,000$ for the public rail. The average fares, average revenue gains per link and passenger volumes per operator are summarized in **Table 4**. In this instance, travelers use 1.6 services on average to reach their destinations for the 4 OD pairs. The base scenario assumes that transfers are costless to users and operators.

### 5.1.4. Government rail acquisition

When the rail lines are acquired by the government, the stable outcome may change from revenue maximization to consumer surplus maximization (buyer-optimal): $\max_{p_{rf}} Z = \sum_{s \in S} u_s(p_{rf})$. This means the stable outcome objective value for the two operators is $Z = \sum_{r \in R_1} p_{r1} z_r + \sum_{s \in S} u_s(p_{r2})$. In this scenario the private operator's revenues increase while those of the public rail service diminish. The problem is that all O-D pairs end up having to use both the public and private operator links to complete their trips. As a result, any reduction in outcome made by the public operator gets absorbed by the private operator instead of going to the travelers. This fact is established in the first term of Eq. (6).

### 5.1.5. Firm entry

What could be the effect of a new operator entering the market on users and on the other transit services? We assume that this new Operator 3 (green in **Figure 7**) is a direct competitor to the services offered by the western rail line (113-112-103-101) and operates parallel links connecting the same nodes with $0 cost transfer links. This new operator only carries 25% of the rail capacities for those links. Suppose this new service also has significantly lower operating costs and travel times (travel times and operating cost set to 25% of the competing service links), which will make it a strong competitor to the rail service.

The results of this analysis are surprising. **Table 4** shows that the base flows among the two operators is 12000 and 9000. When the new operator joins, the 9000 is split to 5000 and 4000 for the new operator, which corresponds to all the demand that was originally using the western line. In other words, the new operator obtains the entire passenger volume of the western line but cannot generate any profit. The effect of strong competition also ends up diminishing the rail line's revenues, while the payoff of the private bus operator is maximized. This can be attributed to the geometry of the network: since all O-D pairs need to use the services of the flexible operator it is completely logical that this will be the only service to benefit from this situation.

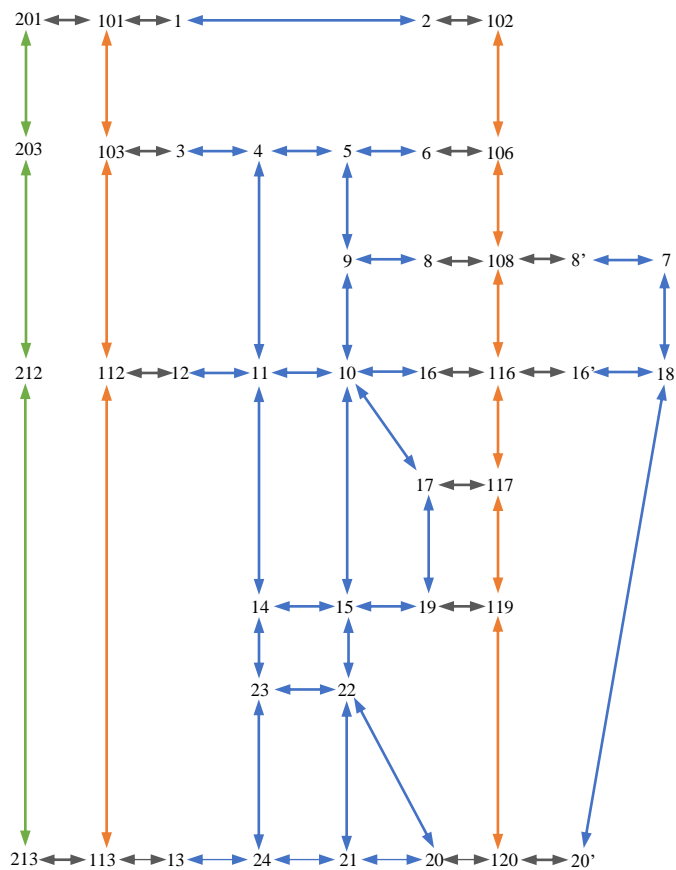

**Figure 7**. Sioux Falls scenario with 3 operators.

### *5.1.6. Capacity effects on MaaS market*

The model allows planners to evaluate the effects of changes in capacity of one operator's link to the performance of all the operators and the users. Such capacity changes may refer to roadway capacity, frequency changes in urban rail lines (effective line capacities), station capacity or queueing operations, fleet size of an on-demand service, or a rebalancing strategy for a bike sharing service that can effectively increase flow capacity along different corridors.

We refer again to our base duopoly scenario in order to quantify the benefits of capacity increases on users and operators. From the matching problem solution described in Eq. (4) we see that link 58 capacity is binding. We test two capacity increases. In this first scenario, we increase capacity of link 58 from 4824 to 4900; in the second case, we go further up to 5000. With a change in capacity, the assignment solution to Eq. (4) may change, and that also impacts the stable outcomes.

While the first scenario does not significantly affect our measures, increasing capacity to 5000 significantly favors the rail operator. The average rail fare increases from $2 to $3, while the private bus operator is forced to reduce its average fare from $2 to $1.3.

*5.1.7. Technological change*

Consider the impact of such technological improvements as new routing or matching algorithms that can reduce operator cost and/or travel disutility on the stable outcome of the MaaS market. We assume that a new fleet management algorithm is deployed by Operator 1 that reduces their links' operating costs by 50% and travel times for users on the same links by 20%. This may represent savings in users' wait times, reduction in detours reducing vehicles miles traveled or empty trips, etc. Under this new technology, the results are quite surprising. While the revenues for Operator 1 increase by 12.5% as expected, we also see the rail service gaining 125% in revenue. The improved performance of the private "feeder" service results in allowing the rail service to respond with higher fares (from $2 to $4.50) and increase revenue distributed from user gains. This fact is attributed to the existing mutualistic relationship (see Chow and Sayarshad, 2014) between the two operators, as no operator alone can provide a complete service to users in the market setting.

**5.2. Sioux Falls example with complete O-D demand and transfer links**

To illustrate the scalability of our algorithm, we solve the Sioux Falls network example using the complete 552-OD demand matrix shown in **Figure 8** to demonstrate the model tractability in comparison to explicit path enumeration. In addition, the transfer links are changed from $0 cost to $2 cost. The link costs and capacities remain the same as shown in **Table 3**.

**Table 5a** reports the matching problem optimum for the complete Sioux Falls test case. Since network characteristics are only intended to demonstrate that our model scales efficiently, we do not make any further calculations about links operated by the rail service (frequencies, travel times etc.). Transfer links are listed as regular links with no operating cost and belong to a dummy operator.

The revenue gains and pricing for operators are shown in Table 5b, which we acknowledge is not guaranteed to be a unique distribution among operators (although their total revenue in the platform should be unique). Passenger utilities of $U_s = 40$ are assumed. The rail operator is to charge a single constant fare for all travelers.

The solution time for the stable outcome problem using **Algorithm 1** is 17 seconds. To compare against the original LP formulation in Eqs. (5) – (6) using explicit path enumeration, we used a modified depth-first search algorithm from the NetworkX Python library, originally proposed by Sedgewick (2001). The solution time for the Sioux Falls network exceeded 2 hours, which is significantly slower.

|    | 1  | 2 | 3 | 4  | 5  | 6 | 7  | 8  | 9  | 10 | 11 | 12 | 13 | 14 | 15 | 16 | 17 | 18 | 19 | 20 | 21 | 22 | 23 | 24 |
|----|----|---|---|----|----|---|----|----|----|----|----|----|----|----|----|----|----|----|----|----|----|----|----|----|
| 1  | 0  | 1 | 1 | 5  | 2  | 3 | 5  | 8  | 5  | 13 | 5  | 2  | 5  | 3  | 5  | 5  | 4  | 1  | 3  | 3  | 1  | 4  | 3  | 1  |
| 2  | 1  | 0 | 1 | 2  | 1  | 4 | 2  | 4  | 2  | 6  | 2  | 1  | 3  | 1  | 1  | 4  | 2  | 0  | 1  | 1  | 0  | 1  | 0  | 0  |
| 3  | 1  | 1 | 0 | 2  | 1  | 3 | 1  | 2  | 1  | 3  | 3  | 2  | 1  | 1  | 1  | 2  | 1  | 0  | 0  | 0  | 0  | 1  | 1  | 0  |
| 4  | 5  | 2 | 2 | 0  | 5  | 4 | 4  | 7  | 7  | 12 | 14 | 6  | 6  | 5  | 5  | 8  | 5  | 1  | 2  | 3  | 2  | 4  | 5  | 2  |
| 5  | 2  | 1 | 1 | 5  | 0  | 2 | 2  | 5  | 8  | 10 | 5  | 2  | 2  | 1  | 2  | 5  | 2  | 0  | 1  | 1  | 1  | 2  | 1  | 0  |
| 6  | 3  | 4 | 3 | 4  | 2  | 0 | 4  | 8  | 4  | 8  | 4  | 2  | 2  | 1  | 2  | 9  | 5  | 1  | 2  | 3  | 1  | 2  | 1  | 1  |
| 7  | 5  | 2 | 1 | 4  | 2  | 4 | 0  | 10 | 6  | 19 | 5  | 7  | 4  | 2  | 5  | 14 | 10 | 2  | 4  | 5  | 2  | 5  | 2  | 1  |
| 8  | 8  | 4 | 2 | 7  | 5  | 8 | 10 | 0  | 8  | 16 | 8  | 6  | 6  | 4  | 6  | 22 | 14 | 3  | 7  | 9  | 4  | 5  | 3  | 2  |
| 9  | 5  | 2 | 1 | 7  | 8  | 4 | 6  | 8  | 0  | 28 | 14 | 6  | 6  | 6  | 9  | 14 | 9  | 2  | 4  | 6  | 3  | 7  | 5  | 2  |
| 10 | 13 | 6 | 3 | 12 | 10 | 8 | 19 | 16 | 28 | 0  | 40 | 20 | 19 | 21 | 40 | 44 | 39 | 7  | 18 | 25 | 12 | 26 | 18 | 8  |
| 11 | 5  | 2 | 3 | 15 | 5  | 4 | 5  | 8  | 14 | 39 | 0  | 14 | 10 | 16 | 14 | 14 | 10 | 1  | 4  | 6  | 4  | 11 | 13 | 6  |
| 12 | 2  | 1 | 2 | 6  | 2  | 2 | 7  | 6  | 6  | 20 | 14 | 0  | 13 | 7  | 7  | 7  | 6  | 2  | 3  | 4  | 3  | 7  | 7  | 5  |
| 13 | 5  | 3 | 1 | 6  | 2  | 2 | 4  | 6  | 6  | 19 | 10 | 13 | 0  | 6  | 7  | 6  | 5  | 1  | 3  | 6  | 6  | 13 | 8  | 8  |
| 14 | 3  | 1 | 1 | 5  | 1  | 1 | 2  | 4  | 6  | 21 | 16 | 7  | 6  | 0  | 13 | 7  | 7  | 1  | 3  | 5  | 4  | 12 | 11 | 4  |
| 15 | 5  | 1 | 1 | 5  | 2  | 2 | 5  | 6  | 10 | 40 | 14 | 7  | 7  | 13 | 0  | 12 | 15 | 2  | 8  | 11 | 8  | 26 | 10 | 4  |
| 16 | 5  | 4 | 2 | 8  | 5  | 9 | 14 | 22 | 14 | 44 | 14 | 7  | 6  | 7  | 12 | 0  | 28 | 5  | 13 | 16 | 6  | 12 | 5  | 3  |
| 17 | 4  | 2 | 1 | 5  | 2  | 5 | 10 | 14 | 9  | 39 | 10 | 6  | 5  | 7  | 15 | 28 | 0  | 6  | 17 | 17 | 6  | 17 | 6  | 3  |
| 18 | 1  | 0 | 0 | 1  | 0  | 1 | 2  | 3  | 2  | 7  | 2  | 2  | 1  | 1  | 2  | 5  | 6  | 0  | 3  | 4  | 1  | 3  | 1  | 0  |
| 19 | 3  | 1 | 0 | 2  | 1  | 2 | 4  | 7  | 4  | 18 | 4  | 3  | 3  | 3  | 8  | 13 | 17 | 3  | 0  | 12 | 4  | 12 | 3  | 1  |
| 20 | 3  | 1 | 0 | 3  | 1  | 3 | 5  | 9  | 6  | 25 | 6  | 5  | 6  | 5  | 11 | 16 | 17 | 4  | 12 | 0  | 12 | 24 | 7  | 4  |
| 21 | 1  | 0 | 0 | 2  | 1  | 1 | 2  | 4  | 3  | 12 | 4  | 3  | 6  | 4  | 8  | 6  | 6  | 1  | 4  | 12 | 0  | 18 | 7  | 5  |
| 22 | 4  | 1 | 1 | 4  | 2  | 2 | 5  | 5  | 7  | 26 | 11 | 7  | 13 | 12 | 26 | 12 | 17 | 3  | 12 | 24 | 18 | 0  | 21 | 11 |
| 23 | 3  | 0 | 1 | 5  | 1  | 1 | 2  | 3  | 5  | 18 | 13 | 7  | 8  | 11 | 10 | 5  | 6  | 1  | 3  | 7  | 7  | 21 | 0  | 7  |
| 24 | 1  | 0 | 0 | 2  | 0  | 1 | 1  | 2  | 2  | 8  | 6  | 5  | 7  | 4  | 4  | 3  | 3  | 0  | 1  | 4  | 5  | 11 | 7  | 0  |

**Figure 8.** O-D demand (x100) for Sioux Falls network

**Table 5a: Link flow assignment results for Sioux Falls network**

| Link | Flow | Link | Flow | Link | Flow |
|---|---|---|---|---|---|
| (20, 21) | 6400 | (23, 22) | 8100 | (19, 15) | 15500 |
| (14, 11) | 15400 | (9, 5) | 11000 | (20, 22) | 10917 |
| (22, 23) | 8100 | (15, 19) | 15500 | (103, 101) | 6000 |
| (2, 1) | 3800 | (10, 16) | 16183 | (11, 14) | 15400 |
| (13, 113)* | 11000 | (117, 17)* | 16200 | (17, 117)* | 16200 |
| (106, 102) | 6600 | (1, 101)* | 6000 | (19, 119)* | 15100 |
| (10, 15) | 16617 | (112, 113) | 10900 | (21, 20) | 6300 |
| (23, 24) | 6200 | (3, 4) | 7200 | (103, 3)* | 5800 |
| (4, 11) | 6900 | (7, 18) | 15900 | (6, 5) | 8800 |
| (102, 106) | 6600 | (22, 20) | 10817 | (102, 2)* | 6600 |
| (18, 20) | 15583 | (1, 2) | 3800 | (16, 18) | 17983 |
| (112, 12)* | 10400 | (108, 116) | 6700 | (24, 23) | 6200 |

| (18, 7) | 15900 | (112, 103) | 6400 | (14, 23) | 6000 |
| --- | --- | --- | --- | --- | --- |
| (17, 10) | 7200 | (18, 16) | 17983 | (23, 14) | 6000 |
| (16, 116)* | 10100 | (119, 117) | 15600 | (120, 20)* | 2900 |
| (8, 108)* | 11500 | (12, 11) | 12200 | (24, 13) | 12100 |
| (15, 22) | 26917 | (16, 10) | 16183 | (13, 24) | 12100 |
| (24, 21) | 14300 | (9, 10) | 22000 | (20, 120)* | 2900 |
| (108, 8)* | 11500 | (116, 117) | 13200 | (116, 16)* | 10100 |
| (106, 6)* | 9600 | (117, 119) | 15600 | (21, 24) | 14400 |
| (116, 108) | 6700 | (11, 12) | 12200 | (119, 120) | 2900 |
| (22, 21) | 12600 | (120, 119) | 2900 | (108, 106) | 12000 |
| (11, 4) | 7000 | (10, 17) | 7200 | (15, 10) | 16817 |
| (22, 15) | 27017 | (101, 103) | 6000 | (101, 1)* | 6000 |
| (5, 6) | 8800 | (21, 22) | 12600 | (103, 112) | 6400 |
| (113, 13)* | 10900 | (4, 3) | 7200 | (5, 4) | 13100 |
| (11, 10) | 18300 | (106, 108) | 12000 | (15, 14) | 11100 |
| (5, 9) | 11000 | (3, 103)* | 5800 | (7, 8) | 10600 |
| (8, 7) | 10600 | (119, 19)* | 15100 | (8, 9) | 3000 |
| (117, 116) | 13200 | (4, 5) | 13100 | (2, 102)* | 6600 |
| (10, 9) | 22100 | (9, 8) | 3000 | (12, 112)* | 10300 |
| (14, 15) | 11100 | (113, 112) | 11000 | (20, 18) | 15483 |
| (23, 22) | 8100 | (6, 106)* | 9600 | (10, 11) | 18500 |

Links with * represent operator transfers

**Table 5b: Pricing and ridership breakdown by operator**

| Operator | Revenue ($) | Avg fare ($) | Min fare ($) | Max fare ($) | Passengers | Operating costs ($) |
| --- | --- | --- | --- | --- | --- | --- |
| Bus Service (1) | 6,509,832 | 23.68 | 1.00 | 38.00 | 274900 | 186 |
| Rail (2) | 217,466 | 1.00 | 1.00 | 1.00 | 217466 | 128 |

### 5.3. Discussion

The computational experiments with Sioux Falls verify that (1) **Algorithm 1** is equivalent to explicit path enumeration for determining the stability constraints; and (2) **Algorithm 1** can be run in 17 seconds while the explicit path enumeration takes longer than 2 hours.

Scenarios related to operator interactions and dependencies can be captured. Pricing and changes in generalized travel costs can also be evaluated, allowing us to quantify the costs of new algorithmic developments not only on its own operator, but also on other operators and travelers in the market. We summarize several key lessons learned.

- Having a firm enter to compete directly with Operator 2 can result in significant advantages to a third party; in this case, Operator 1 can benefit greatly by allowing them to increase their prices.

- Capacity increases even for single links have nonlinear effects, where exceeding some threshold improvement can lead to a significant shift in assignment and stable outcomes, as we see going from 4824 to 4900 to 5000 for link 58. These impacts effect the revenues of other operators as well, so this model allows a platform to monitor the effect of one operator's capacity investments on their competitors.
- Technological changes that impact either system-wide (or operator-wide) operating costs, travel disutilities, and capacities can be evaluated for the whole market. Depending on the type of relationship between the operators, it can be beneficial for multiple parties (if mutualistic) or detrimental to other parties (if parasitic).

## 6. CONCLUSION

With the emergence of MaaS ecosystems, public agencies need modeling tools to consider trade-offs when facilitating markets with private operators. For the most part, such modeling tools do not exist. Recent work from Rasulkhani and Chow (2019) sought to rectify this but only capture line-level interactions between operators and users without allowing users to make multimodal trips. A new modeling framework is proposed using the MCND as a capacitated link-based matching model to fully capture network effects and user paths.

Under this setting, we propose path-based stability conditions and solution algorithms for deriving stable outcomes corresponding to path flows obtained from an MCND along with link capacity dual variables. The modeling framework is tested on an illustrative network as well as in a series of comprehensive experiments on the Sioux Falls network to demonstrate the model's capabilities. As we can see from the lessons learned, the strength of this model lies in using stability conditions to link network design decisions and algorithmic policies (effects on link capacities, operating costs, travel disutilities) as well as market dynamics (firm entry, market consolidation, subsidies, and bundled pricing) to capture market performance for both operators and users.

There are several directions for future research. The non-uniqueness issue suggests there is a problem to determine the minimum perturbations to the paramaters of the MCND such that an empty stable outcome space reaches a unique stable outcome. This is an inverse many-to-many assignment game. The current formulation of stability conditions correspond to planning-level decisions. However, we are also looking into dynamic stability conditions to address day-to-day operations based on probabilistic user route preferences (analogy of stochastic user equilibrium for the MaaS market), flow-dependent travel disutility functions for nonlinear congestion effects and time-dependent demand patterns. Empirical studies using the model and calibrating the model to real data are also important. There are many other fields that the proposed assignment game and algorithm can be applied to beyond MaaS platforms: freight, airlines, other two-sided markets with network effects, and other network flow games where user preferences are of import.

## ACKNOWLEDGEMENTS

This research was conducted with support from NSF CMMI-1634973.

## APPENDIX A: PROOF OF LEMMA 1

*Proof.*
   Operating on the cooperative path (links (1,2) and (2,3)) requires the following condition in Eq. (A1) derived from Eq. (4a).

$$t_{12}d + t_{23}d + c_{12} + c_{23} < t_{13}d + c_{13} \tag{A1}$$

where $d$ is the OD demand for (1,3). The feasibility and stability conditions in this case are shown in Eq. (A2).

$$u + p_{123,b} + p_{123,o} = U - t_{12} - t_{23} \tag{A2a}$$

$$u + p_{123,o} \geq U - t_{13} - c_{13} \tag{A2b}$$

where $U$ is the utility for each travelers and $u$ is the payoff to users. Defining Eq. (13a) in terms of $U$ and substituting it into Eq. (A2B), we get Eq. (A3).

$$t_{13} + c_{13} - t_{12} - t_{23} \geq p_{123,b} \tag{A3}$$

From feasibility condition we know that $p_{123,b}d + p_{123,o}d \geq c_{12} + c_{23}$, so $p_{123,b} \geq \frac{c_{12}+c_{23}}{d} - p_{123,o}$. Substituting the inequality with $p_{123,b}$ into Eq. (14) leads to the following bound for the operator's price in Eq. (A4).

$$p_{123,o} \geq \frac{c_{12} + c_{23}}{d} - t_{13} - c_{13} + t_{12} + t_{23} \tag{A4}$$

Since there's just one price, $p_{123,o}$ simplifies to $p_o$. ∎

**APPENDIX B: PROOF OF LEMMA 2**

*Proof.*
   If link flow $x_{23}^b = 0$, then we get the following feasibility and stability conditions in Eq. (B1).

$$u + p_{2b} + p_{2o} = U - t_{12} - t_{23}^o - t_{34} \tag{B1a}$$

$$u + p_{2b} \geq U - t_{12} - t_{23}^b - t_{34} - c_{23}^b \tag{B1b}$$

Incorporating Eq. (B1a) into Eq. (B1b) results in Eq. (B2).

$$p_{2o} \leq t_{23}^b - t_{23}^o + c_{23}^b \tag{B2}$$

If link flow $x_{23}^b > 0$, the only difference is that $y_{23}^b = 1$, which means $c_{23}^b$ drops out of the inequality, resulting in the final expression. However, the expression $t_{23}^b - t_{23}^o$ is simply 0 since the flow only occurs if $t_{23}^o = t_{23}^b$. ∎